# Luminescence of the $Cs_2ZrCl_6$ under high pressure

L.-I. Bulyk [a], D. Wlodarczyk [a], S.S. Nagorny [b, c], V.V. Nahorna [d], P. Wang [d], and A. Suchocki [a]

[a] Institute of Physics, Polish Academy of Sciences, Al. Lotników 32/46, 02-668, Warsaw, Poland
[b] Department of Physics, Engineering Physics and Astronomy, Queen's University, 64 Bader Lane, K7L 3N6 Kingston, ON, Canada
[c] Arthur B. McDonald Canadian Astroparticle Physics Research Institute, 64 Bader Lane, K7L 3N6 Kingston, ON, Canada
[d] Department of Chemistry, Queen's University, 90 Bader Lane, K7L 3N6 Kingston, ON, Canada

## Abstract

The luminescence and Raman spectra of the $Cs_2ZrCl_6$ crystal in a wide range of pressures were studied in this work for the first time. Luminescence measurements were performed up to 10 GPa, while the Raman spectra were measured up to 20 GPa. The luminescence data revealed a linear blue shift of the emission maximum from about 2.5 eV at ambient pressure to 3.1 eV at 5 GPa and a strong intensity quenching. The indirect-to-direct bandgap transition at about 5 GPa, a phenomenon previously predicted only theoretically, was used to explain the strong quenching of the luminescence. This model was confirmed by fitting luminescence intensity data and analysis of the luminescence decay kinetics, which exhibited a shortening of the pulse decay time with the pressure increase. Raman spectra confirmed the stability of $Cs_2ZrCl_6$ up to 20 GPa and showed no evidence of the pressure-induced structural phase transitions. An energetic scheme of excitonic levels, which takes into account the indirect-to-direct band gasp transition, was proposed to explain the rapid luminescence quenching with increasing pressure.

## Introduction

The rapid advancement of optoelectronic devices has been significantly propelled by the development of all-inorganic metal halide perovskites (MHPs). These materials are notable for their straightforward synthesis, adaptable structures, and exceptional optoelectronic properties [1, 2, 3, 4, 5]. Among the various categories of MHPs, zero-dimensional (0D) MHPs stand out due to their unique structures, consisting of isolated metal halide octahedral units or clusters, which contribute to their luminescent properties, making them appealing for a range of applications [1].

Among the various metal halide perovskites, $Cs_2ZrCl_6$ has garnered significant interest due to its emission properties, non-toxicity, and versatile forms of the material in which it can be synthesized – this sets it apart from many other lead-free perovskites [6, 7, 8]. Over the past three years, research on $Cs_2ZrCl_6$ has increasingly combined theoretical calculations with experimental studies, focusing on its photoluminescence (PL) and scintillation properties. Moreover, the emission maximum of $Cs_2ZrCl_6$ can be modified through Zr-site doping or mixed halide substitution [7].

The excited states in MHPs are predominantly filled by excitons, which can be free excitons (FEs), defect-bounded excitons (DBEs), or self-trapped excitons (STEs). In $Cs_2ZrCl_6$, intrinsic luminescence is related to STEs, which are transient elastic lattice distortions induced by strong electron-phonon coupling upon photoexcitation [1]. These excitons are highly localized within the $[ZrCl_6]^{2-}$ units, resulting in broadband emission with a full width at half maximum (FWHM) of over 100 nm under UV excitation [8]. Energies of all the excitons are obviously smaller than the band gap energy of the material, and the width of the band gap is given differently in different works. In many theoretical works the authors estimate the band gap value to about 3.45 – 4.99 eV, see, for example [9, 10], while the experimental studies give results in the range of 4.7 – 4.86 eV, see works [11, 12, 13]. Also, the theoretical work [11] predicts the decrease of the band gap of $Cs_2ZrCl_6$ with the increase of the pressure applied to the sample. In case of studied in this work material $Cs_2ZrCl_6$, its band gaps edges are formed predominantly by the Zr-d and Cl-p electrons for CB and CB [16, 9], respectively, Pressure application increase the splitting of d-levels and it is responsible for the decrease of the band-gap of this material. Similar behavior was observed for $BaWO_4$:Ce crystals [14]. Also, this is a well-known phenomenon for the materials with the perovskite structure, for example, for $Cs_2AgBiBr_6$ [15] and for $CsPbBr_3$ [16] it was both predicted theoretically and confirmed experimentally. According to these works, the minima of the conduction bands are formed by Ag-d and Pb-s electrons for $Cs_2AgBiBr_6$ and $CsPbBr_3$ respectively, while the maxima of the valence bands are formed by Br-p for both materials. The most crucial is the antibonding interaction between these orbitals, which results in unusual behavior of the band gap with pressure and temperature changes.

Temperature-dependent photoluminescence (PL) measurements reveal strong electron-phonon coupling in $Cs_2ZrCl_6$, characterized by a large Huang-Rhys factor (S > 20). This coupling leads to broad emissions and large Stokes shifts (≥ 60 nm) [8]. In addition, $Cs_2ZrCl_6$ exhibits an unusual dependence of luminescence intensity on temperature. Thermally activated delayed fluorescence (TADF) involves the back-conversion of triplet excitons into singlet states via reverse intersystem crossing (RISC), thereby increasing luminescence efficiency [8]. This process, which requires small singlet-triplet splitting, can be achieved by heavy metal atoms that promote strong spin-orbit coupling.

Pressure can significantly alter the physical shape and properties of these materials [17]. For example, in 2D MHPs, pressure can reduce the bandgap by weakening quantum confinement, making them more suitable for photovoltaic applications [18]. Guo et al. investigated the effect of pressure on $Cs_2ZrCl_6$ using theoretical calculations and found it to be mechanically stable up to 75 GPa. At 0 GPa, $Cs_2ZrCl_6$ has an indirect bandgap with the conduction band minimum (CBM) at the X-point and the valence band maximum (VBM) at the Γ-point. Under ~5 GPa pressure, $Cs_2ZrCl_6$ transitions to a direct bandgap structure with both the CBM and VBM at the Γ-point. This indirect-to-direct bandgap transition is attributed to pressure-induced changes in the charge density distribution [11].

Whilst, $Cs_2ZrCl_6$ has been previously studied for its photoluminescence properties as a function of temperature, the effects of high pressure on its luminescence and structural properties remain largely unexplored in experimental research. This study aims to fill this gap by using high-pressure diamond anvil cell (DAC) techniques to systematically investigate the luminescence and Raman spectra of $Cs_2ZrCl_6$ in a wide pressure and temperature range.

## Experimental techniques

Anhydrous CsCl (99.999%) and $ZrCl_4$ (99.99%) powders were used as starting materials. However, the $ZrCl_4$ powder was subjected to a three-stage purification process prior to the synthesis. At the first stage, the powder was dried at 120°C for 3 hours under vacuum to remove moisture followed by the reduction stage at 300°C for 1 hour in a continuous flow of hydrogen. At the second stage, $ZrCl_4$ powder was sealed in a quartz ampoule and then sublimed at 380°C. At the third stage, the reloaded $ZrCl_4$ powder into a new quartz ampoule was subjected to sublimation at 400°C. CsCl grains and purified $ZrCl_4$ powder were then successively mixed in a stoichiometric ratio, thoroughly ground in a mortar with a pestle, and loaded into a tapered quartz ampoule with an inner diameter of 22 mm.

The $Cs_2ZrCl_6$ single crystalline boule was grown by the vertical Bridgman technique. The sealed ampoule with prepared reagents was gradually heated and maintained at 850°C for 24 hours prior to starting the growth to synthesize the stoichiometric compound and to ensure the melt homogeneity. Then, the first crystal growth was performed at a pulling rate of 35 mm/day and with a temperature gradient of 50°C/cm. After this "fast" growth, the obtained boule was processed to remove all visual particulates of impurities, inclusions, and the first-to-freeze section. The second "slow" growth was performed with a pulling rate of 12 mm/day and a temperature gradient of 25°C/cm. The $Cs_2ZrCl_6$ crystal with a diameter of 21 mm and a length of about 70 mm was obtained, and parts of this boule were used in these studies.

An easyLab diamond anvil cell (DAC) was used for high-pressure measurements of $Cs_2ZrCl_6$. Before the measurements, the sample was cleaved and a piece approximately 20-25 μm thick and about 100 x 100 μm in size was loaded into the cell. The ruby micro-pearls were used as pressure sensors, and the argon gas was used as the pressure-transmitting medium. The hydrostaticity of the pressure was monitored by controlling the width of the luminescence lines of the ruby pressure sensor. In this measurement, the FWHM of the ruby lines exhibited minimal variation throughout the entire pressure range, indicating that hydrostatic conditions were consistently maintained with only minimal gradual deterioration, which is impossible to avoid in high-pressure measurements performed in diamond anvil cells [19].

The luminescence was measured in an Oxford Optistat CF104 cryostat. The luminescence was excited using an EKSPLA pulsed Nd:YAG laser and an NT342 series optical parametric oscillator. The laser operates at a maximum frequency of 20 Hz and generates pulses of approximately 3 ns duration. In this study, the luminescence of the $Cs_2ZrC_6$ crystal was excited at 260 nm. The luminescence was analyzed using an Acton SpectraPro SP-2500 spectrometer from Princeton Instruments, which was equipped with a multi-channel detector head from Hamamatsu Photonics and controlled by an MCD c7557-01 controller.

High-pressure Raman and luminescence measurements were performed up to about 20 GPa in Almax easyLab diamond anvil cell (DAC) using II-as diamonds with 0.45 mm standard design culet. The applied pressure transmitting medium (PTM) was argon. Ruby was used as a pressure gauge. Gaskets with 0.15 mm holes were made from Inconel x750 alloy. Raman measurements were performed on S&I Gmbh Monovista CRS+ Raman Microscope (Olympus XYZ IX71 inverted stage) using 532 nm wavelength obtained from Cobolt Samba 04-01 Nd:YAG laser and 0.75 m Acton Princeton monochromator. A nitrogen-cooled CCD camera with a back-thinned PyLoN system was implemented. The laser was focused on the sample using x5 objective and numerical aperture NA=0.13 in backscattering geometry. The size of the laser spot was about 10 μm. with an average power density of the laser about 0.36 mW/μm$^2$. The spectral resolution was approximately 0.5 cm$^{-1}$. A nitrogen-cooled CCD camera with a back-thinned PyLoN system was implemented as a detector.

## Results and discussion

The absorption spectrum of $Cs_2ZrCl_6$ was measured at ambient conditions on the sample of about 0.23 mm thick. The result is presented in Fig. 1a, where the square root of the absorption coefficient and squared absorption coefficient are fitted with a straight line to determine indirect and direct band gaps, respectively. Obtained values for the band gaps are 4.42 eV for indirect and 4.54 eV for direct band gaps. These results are in relatively good agreement with works [11, 12], and relatively bad agreement with works [9, 13]. In all cases, obtained in this work values of the direct and indirect band gaps are underestimated due to strong absorption under the band gap, and high thickness of the sample – about 0.23 mm.

The temperature dependence of the $Cs_2ZrCl_6$ luminescence was measured at two pressures: ambient and around 1.6 GPa. The results are presented in Fig. 1. In Fig. 1b four spectra are presented: at room temperature and at 5 K for both ambient pressure and 1.6 GPa. In Fig. 1c and d the map of the luminescence change with temperature at ambient pressure (c) and at 1.6 GPa (d). It can be observed from the (c) and (d) figures that the intensity of the luminescence initially increases with increasing temperature, reaches a maximum, and then subsequently decreases. This is not a typical temperature behavior of the luminescence, however, it was previously observed and explained in [8]. The authors proposed that this behavior can be attributed to the presence of higher singlet and lower triplet excitonic states. A higher singlet state has a higher quantum yield and shorter decay constant. Consequently, at low temperatures below 100 K, the lower energy triplet state is populated, resulting in a long decay constant of the luminescence and a reduction in the luminescence intensity. During heating, the population of the more energetic singlet state rises, resulting in the decay constant shortening and an increase in the luminescence intensity. At approximately 200 K, the probability of transitioning to the bottom of the conduction band becomes significant. This results in an enhancing probability of a non-radiative deexcitation, which in turn leads to a decrease in the luminescence intensity. This mechanism is corroborated by the decay kinetics measurements conducted in [8], wherein the authors also established the energy gap between singlet and triplet states to be 69 meV and decay constants of 3.5 and 83.55 μs for triplet and singlet states, respectively. Furthermore, as illustrated in Fig. 1b, the emission spectra exhibit no distinct peaks that can be attributed to transitions from the triplet and singlet states. Instead, a single broad peak is observed. However, the spectra presented in the work [8] also do not reveal a two-band structure. It is related to the close energy positions of both types of excitonic states as well as the broad-band transitions between the excited and the ground states.

Another possible explanation for the increase of the luminescence intensity with the temperature can be a structural phase transition. However, in work [20] authors investigated the crystal structure of $Cs_2ZrCl_6$ at low temperatures, and no phase transition was revealed.

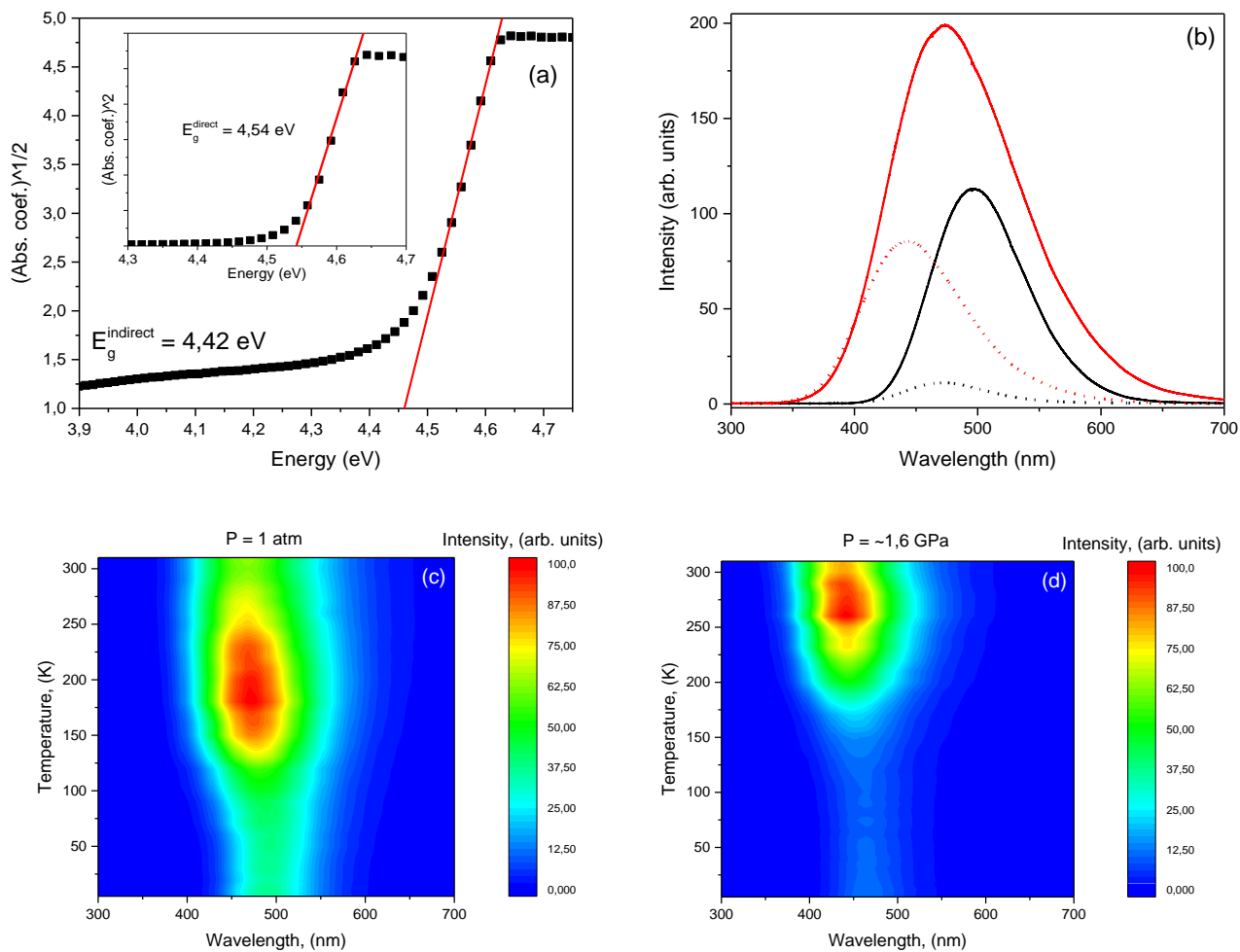

*Fig. 1. The Taucs' plots of the $Cz_2ZrCl_6$ absorption coefficient, measured at ambient conditions, are given in (a). In (b) the luminescence spectra of the $Cs_2ZrCl_6$ are presented. Black curves correspond to 5 K and red curves – 290 K. Ambient pressure spectra are drawn with solid lines, and 1,6 GPa – dotted lines. Temperature dependences of the $Cs_2ZrCl_6$ luminescence at ambient pressure (c) and around 1.6 GPa (d). Luminescence measurements were performed under 260 nm pulsed laser excitation.*

The spectra, on which the maps in Fig. 1 are built, were fitted with Gaussian curves, and the integral intensity as a function of temperature was obtained from these fits. The intensity dependence on temperature is shown in Fig. 2a and Fig. 2b for ambient pressure and 1.6 GPa, respectively. These dependencies were fitted by Equation (1), which assumes an increase in luminescence intensity when the top singlet level of the exciton begins to populate with increasing temperature, and that emission intensity decreases when the population of the bottom of the conduction band increases. Equation (1) postulates a linear pressure dependence of the energetic distance between the low triplet state and the top singlet state, as well as a linear pressure dependence of the energetic distance between the excitonic state and the bottom of the conduction band.

$$I = I_0 \frac{1 + Be^{-\frac{\Delta e_0 + \Delta e_1 P}{kT}}}{1 + e^{-\frac{\Delta e_0 + \Delta e_1 P}{kT}} + Ce^{-\frac{\Delta E_0 + \Delta E_1 P}{kT}}} \quad (1)$$

In this Equation, $\Delta e_0$ and $\Delta E_0$ represent the energy distances from the lowest triplet state to the singlet state and the bottom of the conduction band, respectively. $\Delta e_1$ and $\Delta E_1$ are the pressure coefficients, which identify how the aforementioned energy distances change with pressure. The pressure, temperature, and the Boltzmann constant are assigned as P, T, and k, respectively. The symbols B and C denote the proportionality coefficients, which indicate the probability of the corresponding electronic transitions. Finally, the symbol $I_0$ represents the luminescence intensity at zero temperature.

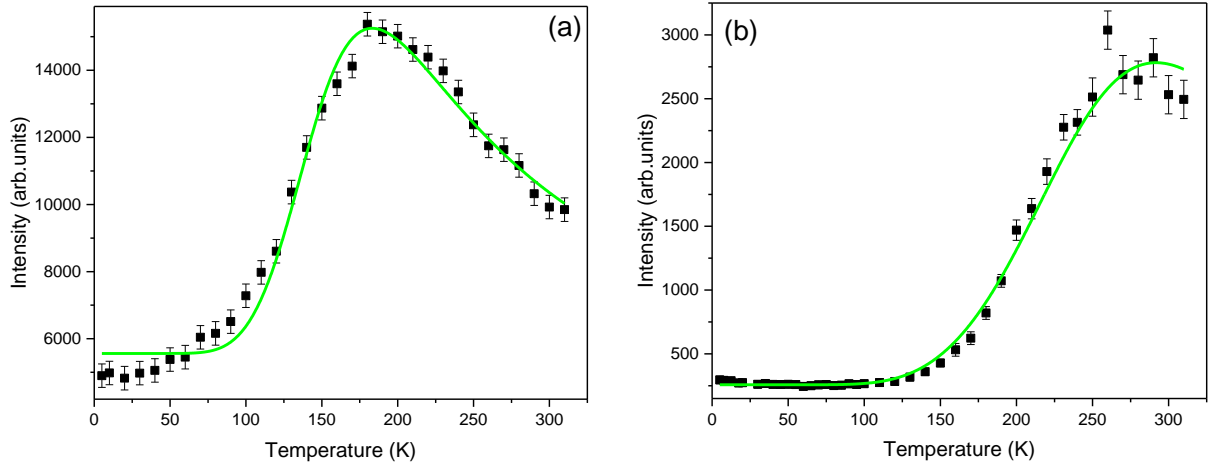

*Fig. 2. Integral intensity of the luminescence as a function of temperature at ambient pressure (a) and at about 1.6 GPa (b). Black squares represent experimental data, and green solid lines represent fit with Equation (1).*

First, the dependence of the intensity on temperature at ambient pressure was fitted (see Fig. 2a). To reduce the number of fit parameters, the energetic distance between singlet and triplet states ($\Delta e_0$) was adopted from the work [8]. Additionally, the pressure (P) and the pressure coefficients ($\Delta e_1$ and $\Delta E_1$) were set to zero. The energy distance from the bottom of the conduction band ($\Delta E_0$), as well as the probability coefficients (B and C), were established from the fit (see the numbers in the first column of Table 1). Subsequently, these values were incorporated into the fitting in Fig. 2b, from which the pressure coefficients ($\Delta e_1$ and $\Delta E_1$) of the energy distances were determined (see the second column of Table 1).

The results of the fittings from the Fig. 2 are in good agreement with the work [8]. However, it is not the only possible explanation for the luminescence intensity dependence on temperature. Another mechanism is described in [20], where authors explain the temperature dependence of the luminescence by the thermally-activated crossover transitions from the trap state to the exciton's excited state, and between the exciton's ground and excited states. The Equation and the values for the fitting parameters were adopted from [20], and compared with the experimental data. The comparison showed, that the fitting is very different from the experimental data, and no good parameters were possible to obtain, which indicates the presence of different deexcitation paths in $Cs_2ZrCl_6$ when excited with the 260 nm laser (used in this work) and 5.5 MeV alpha particles (used in [20]).

*Table 1. Parameters used in Equation (1) were obtained by fitting the luminescence intensity dependence, see Fig. 2 and Fig. 4b.*

|  | Intensity at ambient pressure (Fig. 2a) | Intensity at ~1.6 GPa pressure (Fig. 2b) | Intensity at RT (Fig. 4b, red curve) | Arrhenius fit (Fig. 4b) |
|---|---|---|---|---|
| $I_0$ | 5562±147 | 257.8±7.5 | 0.072±0.013 | 8.8±5.7 |

| $\Delta e_0$ (meV) | 69 (from [8]) | 69 (from [8]) | 69 (from [8]) | -- |
|---|---|---|---|---|
| $\Delta e_1$ (meV) | 0 for amb. press | 7.35±0.53 | ← 7.35 from second fit | -- |
| $\Delta E_0$ (meV) | 97.9±2.3 | ← 97.9 from first fit | ← 97.9 from first fit | 46 (from [11]) |
| $\Delta E_1$ (meV) | 0 for amb. press | 46.2±1.4 | ← 46.2 from second fit | 10 (from [11]) |
| B | 470±46 | ← 470 from first fit | ← 470 from first fit | 28.9±21.7 |
| C | 747±53 | ← 747 from first fit | ← 747 from first fit | -- |
| P (GPa) | 0 for amb. pres | 1.6 GPa | x-axis | x-axis |
| T (K) | x-axis | x-axis | 290 K | 290 K |

The pressure dependence of the luminescence was measured in a series of experiments. For each measurement, another piece of the sample was taken. The series was done to check the repeatability of the results, as well as to check the homogeneity of the sample. The results of the two runs are presented in Fig. 3. As illustrated in the figure, the position of the emission maximum blue-shifts towards higher energies with the pressure increase. The intensity of the luminescence decreases with increasing pressure, reaching almost complete quenching before 10 GPa. The first and second measurements were conducted using argon gas as a pressure-transmitting medium, while during the third measurement, an ethanol-methanol mixture was implemented to enable the attainment of higher pressures.

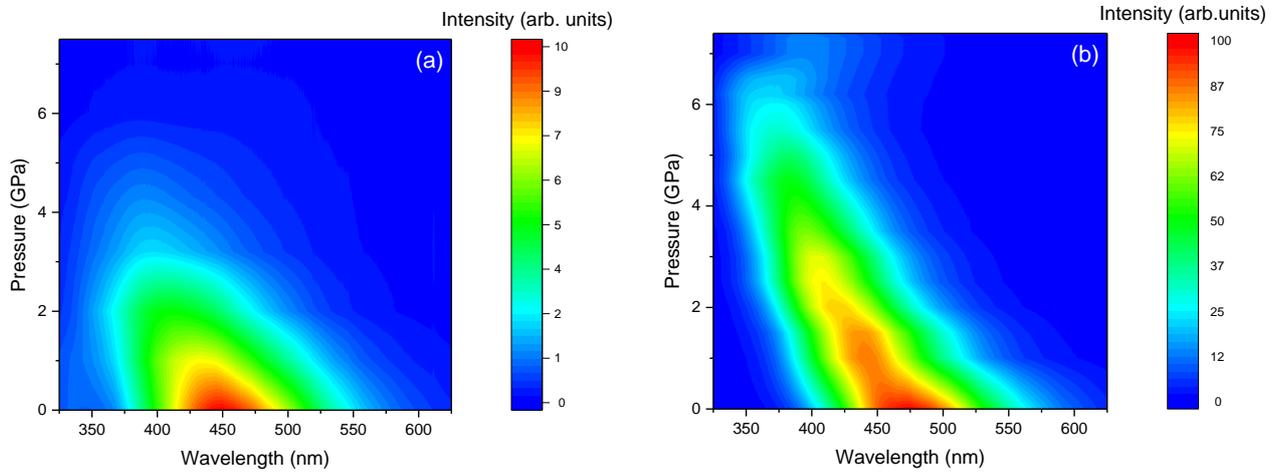

*Fig. 3. The luminescence spectra of $Cs_2ZrCl_6$, measured in two independent runs (first (a) and second (b) measurements) at ambient temperature under high pressure and 260 nm pulsed laser excitation.*

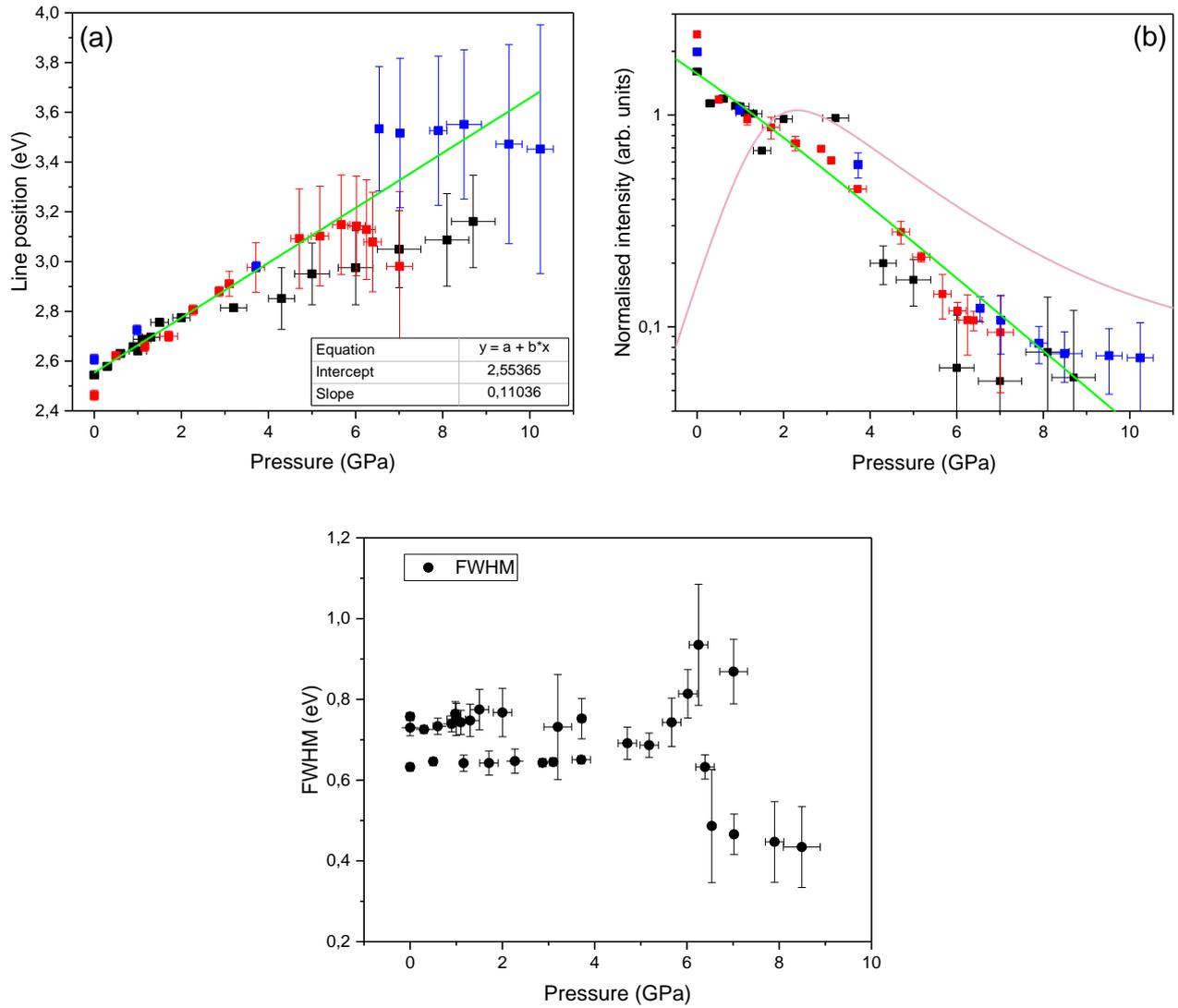

*Fig. 4. Combined data of luminescence position (a) and normalized intensity (b) as a function of pressure at room temperature, with corresponding fits. Black, red, and blue dots represent the first, second, and third measurements, respectively. The green solid line in (a) represents a linear fit, the purple solid line in (b) represents a fit using Equation (1), and the green solid line in (b) represents a fit by the Arrhenius Equation.*

After obtaining the high-pressure luminescence data, the spectra were fitted with a Gaussian function to determine the position and intensity of the luminescence as a function of pressure. At first glance (see Fig. 3), the results of different runs may appear to be inconsistent. However, after examining the fittings (see Fig. 4), it is clear that the deviation in luminescence position at high pressures falls within a high error range. The deterioration of the hydrostatic conditions at high pressures could be a significant contributor to these errors. Henceforth, it can be concluded that the position of the luminescence shifts linearly with increasing pressure, from about 2.5 eV to about 3.1 eV at 5 GPa (about 0.12 eV/GPa). The intensity of the luminescence as a function of pressure, shown on a logarithmic scale in Fig. 4b, decreases by about an order of magnitude when the pressure reaches 5 GPa, confirming the strong quenching with increasing pressure.

Equation (1) was used again for fitting the pressure dependence of the intensity. The values determined in the fits in Fig. 2 (the second column in Table 1) were inserted into Equation (1). However, as can be seen from Fig. 4b (purple curve), the fitting does not follow the experimental data, suggesting that some other mechanism of the luminescence quenching is dominant in this case.

A potential explanation for the observed luminescence quenching is an indirect-to-direct band gap transition, as predicted in theoretical work [11], where it was also determined that the energy difference between the direct and indirect band gaps (46 meV at ambient pressure) changes linearly with the pressure coefficient -10 meV/GPa. These values were incorporated into the modified Arrhenius Equation (2), and the luminescence intensity dependence on pressure was fitted, see Fig. 4b (green line). As can be seen from the figure, the fitting is in good agreement with the experimental data. The parameters of the resulting fit are shown in the last column in Table 1.

$$I = \frac{I_0}{1 + Be^{-\frac{\Delta E_0 + \Delta E_1 P}{kT}}} \quad (2)$$

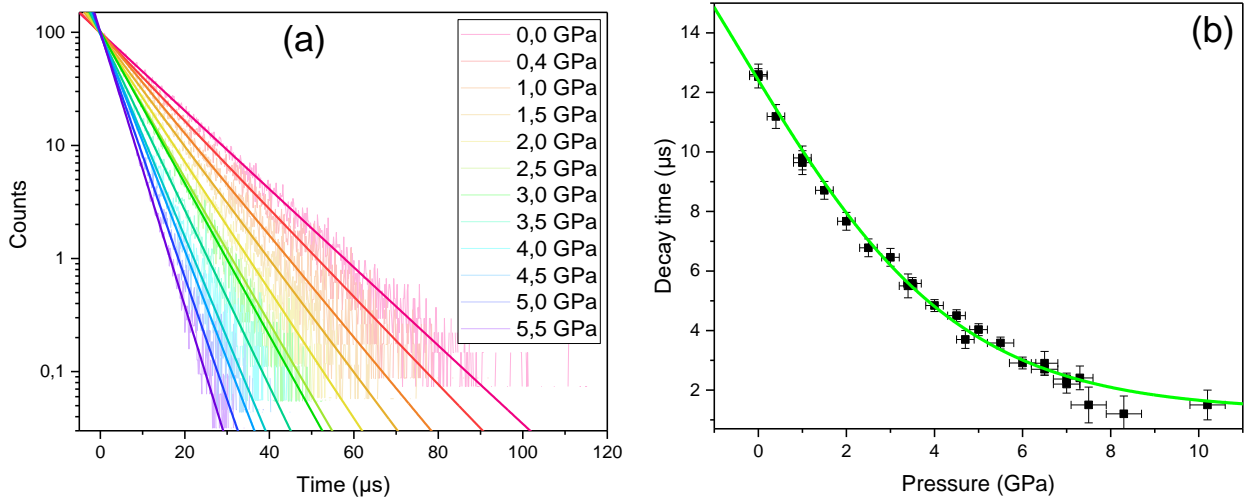

Fig. 5. A few chosen luminescence decay kinetics measured at room temperature (a), and the dependence of the luminescence average decay constant of $Cs_2ZrCl_6$ on pressure at room temperature (b). Black dots represent experimental data, while the green solid line represents fit by Equation (3).

Another possible explanation for the luminescence quenching with the pressure, which could be derived from the fit of experimental data in Fig. 2, is that the energy distance between the singlet and the triplet states in the luminescent $[ZrCl_4]^{2-}$ complex increases during compression. This would result in the depopulation of the more energetic and faster-decaying singlet state, which, as was mentioned above, is also responsible for the more efficient luminescence. Consequently, that may cause the quenching of the luminescence with the pressure increase. However, in this case, the intensity dependence on the pressure should be fitted by Equation (1) with the parameters, determined from fits in Fig. 2, which is impossible (see Fig. 4b, purple curve). Moreover, the average decay constant should increase with the pressure according to this hypothesis. While the average decay constant of the luminescence should decrease with the pressure increase in the case of indirect-to-direct band gap transition.

To make sure which model is correct, the decay kinetics of the $Cs_2ZrCl_6$ luminescence were measured in a wide range of pressures. As can be seen from Fig. 5, the decay constant of the luminescence decreases with the pressure.

The formula for the decay time approximation from [8] was adopted and modified as follows:

$$\tau = \frac{1 + Ae^{-\frac{\Delta E_0 + \Delta E_1 P}{kT}}}{\frac{1}{\tau_r} + \frac{A}{\tau_{nr}}e^{-\frac{\Delta E_0 + \Delta E_1 P}{kT}}} \quad (3)$$

Here, $\tau_r$ and $\tau_{nr}$ represent the decay constants of the radiative (or excitonic) and nonradiative transitions, respectively. $\Delta E_0$ is the energy distance between the indirect and direct band gap. The pressure coefficient, $\Delta E_1$, indicates how the aforementioned energy distance between the indirect and direct band gap changes with pressure. The pressure, temperature, and Boltzmann constant are denoted as P, T, and k, respectively. A is the proportionality coefficient.

*Table 2. Parameters obtained during the fitting of the decay constant dependence on pressure, see Fig. 5b.*

| Parameter | Value (Fig. 5b) |
| --- | --- |
| A | 0.309 ± 0.063 |
| $\Delta E_0$ (meV) | 46 (taken from [11]) |
| $\Delta E_1$ (meV) | -10 (taken from [11]) |
| $\tau_r$ (µs) | 24.6 ± 2.1 |
| $\tau_{nr}$ (µs) | 1.1 ± 0.2 |

The values of $\Delta E_0$ and $\Delta E_1$ were set to 46 meV and -10 meV/GPa, respectively, to align with the data presented in [11] and with the Arrhenius fit in Fig. 4b. The fitting was conducted, and the results are presented in Fig. 5b and Table 2. The value of the radiative (excitonic) decay constant is in good agreement with the one presented in [8]. This provides a strong confirmation that the indirect-to-direct band gap transition is a correct model in the current case, as well, as this evidence is more credible than the fact of not finding a good fit of the experimental data with Equation (1), see Fig. 4b and Table 1.

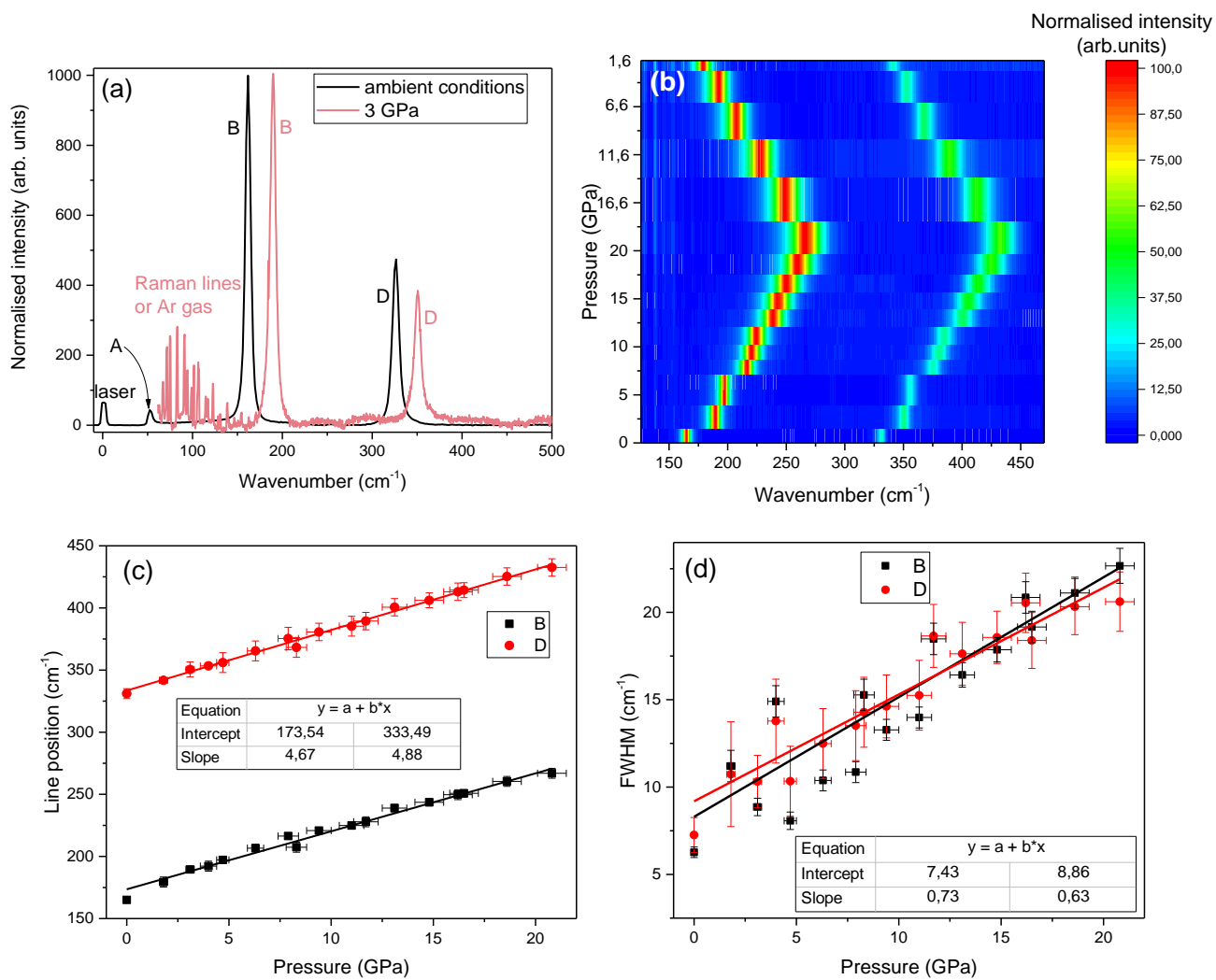

*Fig. 6. Comparison of the room temperature Raman spectra of $Cs_2ZrCl_6$ at ambient pressure and the lowest pressure 3 GPa (a). Map of normalized room temperature Raman spectra on pressure (b). Dependence of the Raman peak position on pressure (room temperature) with the linear fits (c). Dependence of the width (FWHM) of the two strongest Raman peaks on pressure (room temperature) with a linear fit (d).*

It is important to note that the indirect-to-direct band gap phase transition often occurs simultaneously with the structural phase transition. To ascertain whether such a phase transition occurs, Raman spectra of $Cs_2ZrCl_6$ were measured. The energies of the Raman modes of $Cs_2ZrCl_6$ crystal were defined as a function of the hydrostatic pressure (0-20 GPa) at room temperature. The results are presented in Fig. 6. The hydrostatic pressure coefficients of the Raman modes B and D were found to be 4.67 cm$^{-1}$/GPa and 4.88 cm$^{-1}$/GPa, respectively. No evidence of pressure-induced structural phase transitions was observed in this experiment, indicating that the measured crystal is stable up to 20 GPa of hydrostatic pressure. At high pressures, the width of the Raman peaks increases (see Fig. 6b, d), a phenomenon that is typical for high-pressure Raman measurements [21]. The observed increase in width can be attributed to the small deterioration of the hydrostatic conditions. The pressure-dependent Raman spectra are shown in the map in Fig. 6b in the wavenumber range from 130 to 470 cm$^{-1}$. The region below 130 cm$^{-1}$ is omitted in high-pressure measurements because Raman line A of the $Cs_2ZrCl_6$ is covered by the strong Raman signal of the Ar gas, which was used as a pressure transmitting medium. To illustrate the overlap of the A-line with the Ar gas signal, a comparison of the Raman spectra at ambient conditions and the lowest pressure of 3 GPa is shown in Fig. 6a.

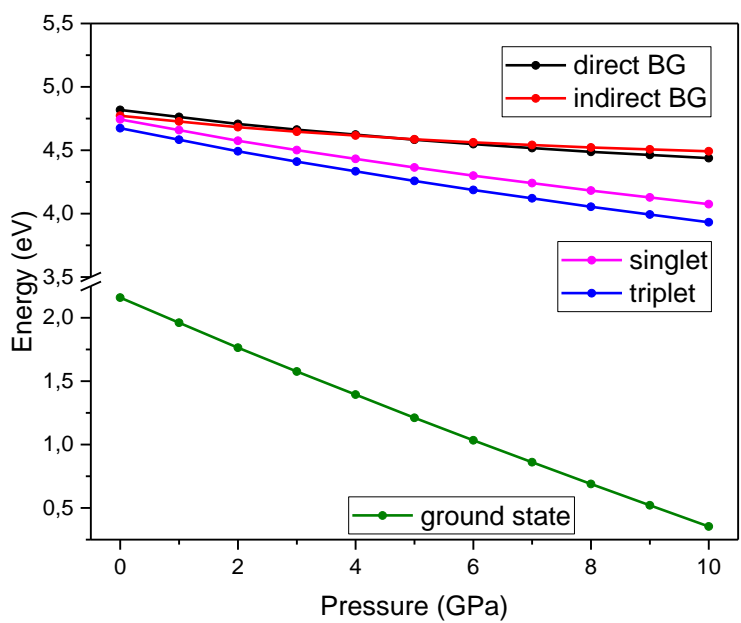

*Fig. 7. Dependence on the pressure of the energy levels of $Cs_2ZrCl_6$ crystal. The data of the indirect (red) and the direct (black) band gap was taken from [11]. The position of the singlet (purple) and triplet (blue) states of the exciton was calculated relative to the band gap using the data obtained during the fits in Fig. 2, Table 1. The ground state (green) of the exciton was calculated using the results of the fit in Fig. 4a, Table 1. The top of the valence band is taken as a reference level.*

The data in Table 1, obtained from the fitting procedures, enables the construction of an energetic scheme representing the $Cs_2ZrCl_6$ exciton behavior as a function of pressure. The indirect and direct band gaps were taken from the aforementioned work [11]. As the calculated band gaps are underestimated, a constant of 1.07 eV was added to the theoretical values to obtain the indirect band gap at zero pressure, which is equal to 4.77 eV, the experimentally established band gap of $Cs_2ZrCl_6$, as cited in [11]. The fitting parameters $\Delta e_0$ and $\Delta E_0$, and $\Delta e_1$ and $\Delta E_1$ from Table 1 were used to calculate the position of the singlet and triplet states of the $Cs_2ZrCl_6$ exciton. The later state was calculated by subtracting $(97.9 + 46.2 * P)$ meV from the indirect band gap (numeric values are taken from Table 1, where the parameters of the fits from Fig. 2 are presented). Afterwards, the position of the singlet state was determined by adding $(69 + 7.35 * P)$ meV to the triplet state (numeric values are taken as well from Table 1, Fig. 2). Finally, the energy of the luminescence was subtracted from the average of the singlet and triplet states to obtain the position of the ground state of the exciton. The energetic scheme of $Cs_2ZrCl_6$ was built after these calculations, see Fig. 7.

The observed pressure-induced increase of energy of exciton luminescence is the most probably related to the decrease of the Huang-Rhys parameter and related decrease of the Stokes shift. Pressure application decrease of the electron-phonon coupling and shifts the parabolas of the excited states of excitons towards more central position of the minima of the excited and the ground states.

## Conclusions

This article exhibits comprehensive spectroscopic studies of the $Cs_2ZrCl_6$ single crystals, yielding substantial insights into the material's properties under high pressure. The findings indicate that at a given constant pressure, the luminescence of the crystal is primarily influenced by the redistribution of an exciton population between the singlet and triplet states, as a function of temperature.

A novel energetic scheme for the excitonic levels in $Cs_2ZrCl_6$ crystal was proposed, which explained the acute luminescence quenching due to the indirect-to-direct band gap transition around 5 GPa. A prompt decrease of the luminescence decay constant with the pressure increase, which supports the proposed energetic scheme, could be described by the presence of two types of transitions – radiative (with 24.6 μs decay constant) and non-radiative (with 1.1 μs decay constant). Moreover, the luminescence emission maximum underwent a blue shift from approximately 2.5 eV to 3.1 eV at 5 GPa, accompanied by a strong quenching during compression.

The indirect-to-direct band gap transition undergoes with no structural phase transition, as evidenced by Raman spectra measurements performed up to 20 GPa. The Raman spectra contain two main peaks located at about 173 $cm^{-1}$ and 333 $cm^{-1}$ at ambient pressure, and they linearly shift to about 220 $cm^{-1}$ and 382 $cm^{-1}$ when the pressure increases up to 10 GPa. This indicates the stability of $Cs_2ZrCl_6$ crystal at these hydrostatic conditions.

All these findings contribute to the understanding of the optoelectronic and structural properties of $Cs_2ZrCl_6$ crystals under high pressure.

Acknowledgments: This work was partially supported by the Polish National Science Centre program SHENG2 of Poland-China cooperation, project number: 2021/40/Q/ST5/00336.